\documentclass[12pt]{article}
\usepackage{alltt}
\usepackage{epsfig}
\usepackage{amsmath}
\usepackage{amssymb}
\usepackage{amsfonts}
\usepackage{times}

\author{A. Gavrielides,  V. Kovanis and P.M. Varangis\\
Nonlinear Optics Center, Phillips Laboratory, \\
Kirtland AFB, NM 87117-5776
\and
T. Erneux and G. Lythe\thanks{Present Address: Center for Nonlinear Studies, Los Alamos National Laboratory, Los Alamos, NM 87545 }\\
Universit\'e Libre de Bruxelles, Optique  Nonlin\'eaire Th\'eorique,\\
Campus Plaine, C.P. 231, 1050 Bruxelles, Belgium}
\title{Coexisting periodic attractors in injection locked diode lasers
}
\date{ 
}

\begin{document}

\maketitle
\begin{abstract}
We present experimental evidence of coexisting periodic attractors in a
semiconductor laser subject to external optical injection. The coexisting
attractors appear after the semiconductor laser has undergone a Hopf
bifurcation from the locked steady state. We consider the single mode rate
equations and derive a third order differential equation for the phase of
the laser field. We then analyze the bifurcation diagram of the time
periodic states in terms of the frequency detuning and the injection rate
and show the existence of multiple periodic attractors.

\newpage\ 
\end{abstract}

\section{Introduction}

Semiconductor lasers have a wide range of applications because they are of
relatively small size, they can be massively produced at low cost, and they
are easy to operate. Applications of semiconductor lasers appear in many
areas such as optical communication and high speed modulation and detection.
Despite their successful technology, semiconductor lasers are quite
sensitive to any external perturbation which may destabilize their normal
output. A small amount of optical feedback resulting from the reflection
from an optical disk or from the end of an optical fiber is sufficient to
generate pulsating instabilities. These oscillations are typically
accompanied by higher intensity of frequency noise and affect the normal
efficiency of the laser.

Systematic experimental studies of semiconductor lasers, in particular long
time series analysis of the intensity are not available because the time
scale of the intensity pulsations is typically in the picosecond regime.
Fourier spectra measurements show a gradual increase of oscillatory
instabilities as parameters are changed but do not reveal what are the
bifurcation mechanisms. Most of the progress on our understanding of these
bifurcations comes from extensive numerical studies of simple models and
their comparison to the experimentally obtained Fourier spectra.
Specifically, these are the Lang and Kobayashi equations \cite{lk} for a
single mode laser subject to optical feedback \cite{mork}, the equations for
two coupled lasers \cite{winful}, and the equations for a laser subject to
external optical injection \cite{lang}. The latter is presumably the
simplest system which allows a study of instabilities induced by an external
perturbation.

The rate equations in normalized form for this system consist of two
equations for the complex electrical field $E$ and the excess carrier number 
$N$ given by: 
\begin{equation}
\label{1}E^{\prime }=(1-ib)NE+\eta E_i, 
\end{equation}
\begin{equation}
\label{2}TN^{\prime }=P-N-P(1+2N)|E|^2 
\end{equation}
where prime means differentiation with respect to time $t$ measured in units
of the photon lifetime $\tau _p$. The term $\eta $$E_i=\eta \exp (-i\Omega
t) $ models the electrical field of the injected signal which is controlled
by changing either its amplitude $\eta $ or its detuning $\Omega $$.$ The
equations are the same for a laser subject to optical feedback or for each
laser in arrays of coupled lasers. They differ only by the last term in Eq. (%
\ref{1}). The fixed parameters $b$, $T$ and $P$ are defined as follows: $b$
is the linewidth enhancement factor ($b=3-6$) which measures the amount of
amplitude-phase coupling; $T\equiv \tau _s/\tau _p$ is the ratio of the
carrier and photon lifetimes ($T\simeq 10^3)$ and $P$ is the dimensionless
pumping current above threshold ($\left| P\right| $ $<1).$ A series of
recent experiments have explored a large variety of dynamical instabilities
at high injection levels and have been simulated successfully by using Eqs. (%
\ref{1}) and (\ref{2}). The comparison is particularly remarkable knowing
the high simplicity of these equations. They correctly describe the observed
period doubling cascade to chaos \cite{p1}, the period 2 bubbles \cite{p2},
as well as the details of the experimental map of all the instability
regions in the detuning versus injection amplitude parameter space \cite{p3}.

Of particular interest are the conditions for locking which corresponds to a
steady phase of the laser field and occurs if the detuning is sufficiently
low. This phenomenon is best analyzed by introducing $E=R\exp [i(\Phi
-\Omega t)]$ into Eqs. (\ref{1}) and (\ref{2}) and by studying the
conditions for a steady phase $\Phi .$ If the injection level is very weak,
it is reasonable to assume that the laser amplitude and the carrier number
will approach their steady state values in the absence of injection (i.e., $%
R\simeq 1$ and $N\simeq 0)$ and that the long time behavior of the laser is
first described by the phase $\Phi .$ From the equations for $R$, $\Phi $
and $N$, it is then straightforward to find that $\Phi $ satisfies Adler
equation 
\begin{equation}
\label{3}\Phi ^{\prime }=\Omega -\eta \sin (\Phi ). 
\end{equation}
Locking now implies the condition $|$$\Omega |\leq \eta .$ In addition to
locking, Eq. (\ref{3}) has been used to investigate the case of an unbounded
phase (i.e., if $|$$\Omega |>$$\eta ).$ Specifically, if $\eta <<\Omega ,$ $%
\Phi \simeq \Omega t$ which implies a small amplitude periodic modulation of
the laser amplitude equation i.e., $R^{\prime }=RN+\eta \cos (\Phi )\simeq
RN+\eta \cos (\Omega t).$

This case has been studied analytically \cite{lenstra} and is called four
wave mixing because of the typical Fourier spectra. However, the practical
interest of (\ref{3}) is limited by the fact that $T$ is a large
parameter. The effect of $T$ large can be first analyzed from Eqs. (\ref{1})
and (\ref{2}) with $\eta =0.$ We find that small perturbations from the
steady state $(R,N)=(1,0)$ slowly decays on an $O(T)$ time interval with
rapid oscillations characterized by period $2\pi /\omega =O(\sqrt{T})$ where 
$\omega \equiv \sqrt{2P/T}$ is the laser relaxation oscillations frequency.
We take this property into account by introducing the new time $s\equiv
\omega t$ and by reformulating Eqs. (\ref{1}) and (\ref{2}) in terms of
normalized deviations from the steady state, namely, $x\equiv N\omega ^{-1}$
and $y\equiv R-1.$ By then investigating the limit $\omega \rightarrow 0$
(equivalently, $T$ large) of the resulting equations for $x$, $y$ and $\Phi
, $ we reobtain (\ref{3}) provided that $\Omega =O(\eta )$ and 
\begin{equation}
\label{4}\eta \leq \omega ^2=O(T^{-1}). 
\end{equation}
Thus, (\ref{3}) is only valid for relatively low injection. It explains
the locking phenomenon and the entrainement of the laser in the four wave
mixing regime but cannot lead to the multiple instabilities observed
experimentally and numerically.

In particular, we show in Figures 1a and 1b experimentally obtained spectra of
a semiconductor laser subject to injection operating at very low detunings
between the frequencies of the slave and master laser. The spectra were
taken at constant injection and detuning. They show the characteristic
sidebands at the relaxation oscillations frequency indicating that a Hopf
bifurcation has taken place from the locked steady state. The laser
exhibited the two spectra at various times. The transition from one spectrum
to another was abrupt strongly suggesting that the laser was visiting two
different periodic states assisted by the intrinsic noise in the system.
This implies that another stable periodic state may coexist with the Hopf
bifurcation periodic state for the same values of the parameters. 
Figures 2a
and 2b show spectra obtained from solving numerically the single mode rate
equations (\ref{1}) and (\ref{2}). The parameters used for these simulations
were $T=1000,$ $b=4,$ $P=1.0,\eta =0.002$ and $\Omega /\omega =0.1$. These
values of the parameters are of the same order as the values that have been
used before by a number of authors to model the HLP-1400 semiconductor
lasers (emitting at $830$ nm). These lasers were also used in this set of
experiments \cite{li}. Gain saturation, known to shift bifurcation points,
did not alter our conclusions, producing only minor modifications of the
spectra. {\it \ }Figure 3 shows the bifurcation diagram of the periodic
states. In addition to the periodic states emerging from the Hopf
bifurcation, there is an isolated branch of periodic states emerging from a
limit point. The spectra shown in Figures 2a and 2b were computed at the
points denoted by A and B in Figure 3, respectively. They compare well with
the experimental spectra shown in Figures 1a and 1b.

Bifurcation diagrams exhibiting coexisting bifurcation and isolated branches
of periodic attractors are not common for nonlinear bifurcation problems.
Similar diagrams have been found numerically for the laser subject to
optical feedback \cite{al} suggesting that the emergence of coexisting
branches of periodic attractors is a general feature of semiconductor laser
instabilities. This motivates analytical studies of the laser rate equations
which we describe in the next two sections. In Section 2, we formulate a
third order phase equation which is a major simplification of the original
laser equations. In Section 3, we analyze this equation and determine
approximations for the steady and time-periodic solutions.

\section{Formulation}

Intensive numerical studies of the first period doubling bifurcation in the
simplest case of zero detuning \cite{te} showed that the linewidth
enhancement factor plays a key role in the problem. Its relatively large
value compared to the other constants in the problem explains why the
intensity oscillations are nearly harmonic in time and how a strong coupling
between phase and laser amplitude is possible. This can be demonstrated by a
new asymptotic analysis of the original equations using $b$ as the order
parameter. To this end, we first introduce the decomposition $E=R\exp
[i(\Phi -\Omega t)]$ into Eqs. (\ref{1}) and (\ref{2}) and reformulate the
resulting three equations in terms of new variables defined by 
\begin{equation}
\label{4a}s\equiv \omega t,\rm{ }x\equiv Nb\omega ^{-1}\rm{ and }y\equiv
b(R-1).
\end{equation}
In (\ref{4a}), time is scaled by the relaxation oscillations frequency $%
\omega \equiv \sqrt{2P/T}$ and $x$, $y$ represent deviations from the zero
injection steady state solution $(R,N)=(1,0).$ The resulting equations for $%
x,$ $y$ and $\Phi $ are then analyzed for large $b$ assuming $\omega
=O(b^{-1})$ (see Appendix). The reduced problem - accurate to $O(b^{-1})$ -
is described in terms of the phase $\Phi $ and is given by (\ref{a9}),
or equivalently, 
\begin{equation}
\label{4p}
\begin{array}{c}
\Phi ^{\prime \prime \prime }+(\xi -2b^{-1}\Delta )\Phi ^{\prime \prime
}+\Phi ^{\prime }-\Delta -\Lambda \cos (\Phi ) \\ 
=b^{-1}\left[ \Lambda \sin (\Phi )\Phi ^{\prime 2}-\Lambda \sin (\Phi
)-2\Phi ^{\prime }\Phi ^{\prime \prime }\right] 
\end{array}
\end{equation}
where prime now means differentiation with respect to time $s.$ The
parameters $\Lambda ,$ $\Delta $ and $\xi $ are defined by 
\begin{equation}
\label{5}\Lambda \equiv \frac{\eta b}\omega ,\rm{ }\Delta \equiv \frac
\Omega \omega \ \ \rm{and}\ \ \xi \equiv \omega \frac{1+2P}{2P}
\end{equation}
and correspond to the normalized injection, normalized detuning and laser
damping parameters, respectively. The leading order problem as $%
b^{-1}\rightarrow 0$ ($\omega $ small but fixed) is

\begin{equation}
\label{6}\Phi ^{\prime \prime \prime }+\xi \Phi ^{\prime \prime }+\Phi
-\Delta -\Lambda \cos (\Phi )=0
\end{equation}
and was previously examined for the case $\Delta =0$ (zero detuning) \cite
{te}. In particular, its validity has been carefully investigated by
comparing the numerical bifurcation diagrams obtained from Eqs. (\ref{1})
and (\ref{2}) and from Eq. (\ref{6}). The two bifurcation diagrams have been
compared for various ranges of values of the parameters $T$ and $b$
including the experimental parameters estimated in \cite{p1}, \cite{p2} and 
\cite{p3}. The agreement between exact and approximate diagrams is excellent
provided that the injection level is not too large (i.e., $\Lambda $ must be 
$O(1)$ or smaller). For the case $\Delta \neq 0,$ a direct comparison
between experimental data and solutions of Eq. (\ref{6}) has been undertaken
if $\Delta $ is close to $2$ and in the four wave mixing region. If $\Delta $
$\simeq 2,$ the laser exhibits two period doubling bifurcations which are
not observed for other values of $\Delta $ and motivated the interest for
this particular case. The optical spectra have been obtained for $\Delta $
below and above $2$ which allow us to study both the sudden transition to
subharmonic resonance and the progressive shift of the frequencies \cite{ga}.

In this paper, we investigate Eq. (\ref{4p}). The steady states and their
linear stability properties are easily determined for small $b^{-1}$ and $%
\omega =O(b^{-1}).$ We find that the steady states emerge from limit points
located at 
\begin{equation}
\label{6a}\Lambda =\Delta 
\end{equation}
and that they may change stability at Hopf bifurcation points located at 
\begin{equation}
\label{6b}\Lambda =\sqrt{\Delta ^2+(\xi -2b^{-1}\Delta )^2+O(b^{-2})}.
\end{equation}
From (\ref{6a}) and (\ref{6b}), we note that a limit point and a Hopf
bifurcation point may coalesce if $\Delta =\Delta _c$ and $\Lambda =\Lambda
_c$ where 
\begin{equation}
\label{new7}\Delta _c\simeq \xi b/2\rm{ and }\Lambda _c=\Delta _c.
\end{equation}
This degenerate Hopf bifurcation point was already noticed from the exact
linear stability analysis \cite{stability}. It corresponds to a zero
eigenvalue and a pair of imaginary eigenvalues of the linearized theory and
may lead to a secondary bifurcation to quasiperiodic oscillations \cite{gh}.

\section{Bifurcation equations}

The phase equation (\ref{6}) is a major simplification of the original
laser equations. It models the laser problem as a weakly damped harmonic
oscillator strongly driven by the phase of the laser field but it is still
too complicated for exact solutions. In this section, we determine an
asymptotic solution of Eq. (\ref{4p}) valid for small $b^{-1},$ $\Lambda ,$ $%
\Delta $ and $\xi $ by the method of multiple scales \cite{kevorkian}$.$ To
this end, we expand the parameters as

\begin{equation}
\label{new9} 
\begin{array}{c}
\qquad \Lambda =b^{-1}(\lambda +b^{-1}\lambda _1+...), 
\rm{ \qquad }\Delta =b^{-1}(\delta +b^{-1}\delta _1+...),\rm{ } \\ \xi
\equiv \omega \frac{1+2P}{2P}=b^{-1}(\sigma +b^{-1}\sigma _1+...) 
\end{array}
\end{equation}
and seek a solution of the form

\begin{equation}
\label{new10}\Phi (s,\tau ,b^{-1})=\Phi _0(s,\tau )+b^{-1}\Phi _1(s,\tau
)+... 
\end{equation}
where $\tau \equiv b^{-1}s$ is defined as a slow time variable. The fact
that the solution depends on two independent time variables implies the
chain rule $\Phi ^{\prime }=\Phi _s+b^{-1}\Phi _\tau $ where subscripts mean
partial derivatives.

Inserting (\ref{new9}) and (\ref{new10}) into Eq. (\ref{4p}) and equating to
zero the coefficients of each power of $\ b^{-1}$ leads to a sequence of
linear problems for the coefficients $\Phi _0,\Phi _1,...$ Solving the
equation for $\Phi _0$ gives 
\begin{equation}
\label{7}\Phi =A\sin (s+v)+B+O(b^{-1}) 
\end{equation}
where the amplitudes $A$ and $B$ and the phase $v$ are functions of the slow
time $\tau .$ We obtain equations for $A$, $B$ and $v$ by considering the
problem for $\Phi _1$ and by applying solvability conditions \cite{kevorkian}%
. The equation for $v$ is simply $v^{\prime }=0$ which means that $v$ is
equal to its initial value. The equations for $A$ and $B$ are given by 
\begin{equation}
\label{8}A^{\prime }=-\frac 12\sigma A+\lambda \sin (B)J_1(A) 
\end{equation}
\begin{equation}
\label{9}B^{\prime }=\delta +\lambda \cos (B)J_0(A) 
\end{equation}
where $J_0(A)$ and $J_1(A)$ are Bessel functions which come from the
expansion of $\cos (\Phi )$ in Fourier series. Note that the modified phase
equation (\ref{4p}) and the leading order phase equation (\ref{6}) lead to
the same amplitude equations at this order of the analysis. Eqs.(\ref{8})
and (\ref{9}) admit steady state solutions which we analyze in detail. The
simplest solution is given by 
\begin{equation}
\label{9a}(1)\rm{ }A=0\quad{\rm and }\quad
\Lambda =-\Delta \cos {}^{-1}(B)>0 
\end{equation}
and corresponds to the steady state of the original equation (\ref{6}). The
second steady state solution of Eqs.(\ref{8}) and (\ref{9}) is characterized
by the fact that $A\neq 0$ and corresponds to time-periodic solutions of Eq.
(\ref{6}). It is instructive to examine this solution for $\Delta =0$ first$%
. $ From Eq. (\ref{9}) with $B^{\prime }=\delta =0,$ we find two
possibilities. The first family of solutions verifies the condition cos$%
(B)=0 $ and is described by 
\begin{equation}
\label{10}(2a)\rm{ }B=\pm \frac \pi 2\quad{\rm and }\quad
\Lambda =\pm \frac{\xi A}{%
2J_1(A)}>0. 
\end{equation}
From (\ref{10}), we determine a Hopf bifurcation branch emerging from
Solution $(1)$ at ($\Lambda ,A,B)=(\Lambda ^H,0,\frac \pi 2).$ $\Lambda ^H$
is found from (\ref{10}) by taking the limit $A\rightarrow 0.$ We obtain 
\begin{equation}
\label{11}\Lambda ^H=\xi 
\end{equation}
which is correctly matching (\ref{6b}) if $\Delta =0.$ But Eq. (\ref{10})
describes other branches of solutions. These branches of solutions are
isolated and emerge from limit points that satisfy the condition $\Lambda
^{\prime }(A)=0.$ Using the second expression in (\ref{10}), we find that
these limit points are located at $(\Lambda ,A,B)=(\Lambda ^L,A^L,\pm \frac
\pi 2)$ where $A^L$ and $\Lambda ^{L\rm{ }}$ satisfies the condition 
\begin{equation}
\label{12}J_2(A^L)=0\quad{\rm and }\quad\Lambda ^L=\pm \frac{\xi A^L}{2J_1(A^L)}>0. 
\end{equation}

The second family of solutions characterized by an $A\neq 0$ verifies the
condition $J_0(A)=0$ and is described by 
\begin{equation}
\label{13}(2b)\rm{ }J_0(A)=0\quad{\rm and }\quad
\Lambda =\frac{\xi A}{2\sin
(B)J_1(A)}>0.
\end{equation}
This solution emerges from Solution $(2a)$ at the pitchfork bifurcation
point located at ($\Lambda ,A,B)=(\Lambda ^{PF},A^{PF},\pm \frac \pi 2)$
where 
\begin{equation}
\label{13a}J_0(A^{PF})=0\quad{\rm and }\quad
\Lambda ^{PF}=\pm \frac{\xi A^{PF}}{%
2J_1(A^{PF})}>0.
\end{equation}

In the case of nonzero detuning ($\Delta \neq 0),$ we find that all $A\neq 0$
steady state solutions of Eqs. (\ref{8}) and (\ref{9}) are described by (in
implicit form) 
\begin{equation}
\label{17}\rm{ }\Lambda =\left[ \left( \frac \Delta {J_0(A)}\right)
^2+\left( \frac{\xi A}{2J_1(A)}\right) ^2\right] ^{1/2}
\quad{\rm and }\quad B=\arccos
\left( -\frac \Delta {\Lambda J_0(A)}\right) . 
\end{equation}
Eq. (\ref{17}) reveals several branches of solutions which converge to
Solution $(2a)$ and Solution $(2b)$ as $\Delta \rightarrow 0$ . In Figure 4,
we show the bifurcation diagram of the solutions of Eqs. (\ref{8}) and (\ref
{9}) in terms of $A$ vs $\Lambda .$ The solid and dotted lines correspond to 
$\Delta =0$ and $\Delta =0.01,$ respectively. All stable and unstable
branches of solutions have been drawn. Note the coexistence of multiple
branches of solutions emerging from limit points. If $\Delta =0,$ these
limits points correspond to simple roots of Bessel functions (see Eq.(\ref
{12}) and Eq.(\ref{13a})).

The numerical bifurcation diagram of the stable solutions of Eq.(\ref{6})
with $\Delta =0$ is shown in Figure 5. Figures 5a and 5b represent the
extrema of the phase $\Phi $ and the intensity variable $y\equiv b(R-1),$
respectively. The pitchfork bifurcation at (\ref{13a}) leading to two
distinct branches of $B$ is clearly identified in Figure 5a. These two
branches are not seen in Figure 5b because $y=\Phi ^{\prime \prime }\simeq
-A\sin (s+\upsilon )$ from (\ref{a7a}) and is independent of $B$, in first
approximation (equivalently, the extrema of $y$ are equal to $\pm $ $A,$ in
first approximation). The figure shows one isolated branch of stable
periodic states that coexist with the bifurcation branch. The bifurcation at 
$\Lambda \sim 0.65$ is a period doubling bifurcation which is not predicted
by Eqs. (\ref{8}) and (\ref{9}) and which requires a different analysis \cite
{te}.

The unfolding of the $\Delta =0$ bifurcation diagram as $\Delta $ is
progressively increased is illustrated by the dotted lines in Figure 4. In
Figure 6, we represent the extrema of $y$ for all stable solutions determined
numerically from (\ref{6}) with $\Delta =0.01$. The agreement between
numerical and analytical solutions is excellent (compare Figure 4 and Figure 6).

The amplitude equations (\ref{8}) and (\ref{9}) admit a third family of
solution if $|\Delta /\Lambda |>1.$ These solutions correspond to $A=0$ and
an unbounded $B$ satisfying Adler's equation, $B^{\prime }=\Delta +\Lambda
\cos (B).$ We have verified that all solutions of this equations are
linearly stable with respect to small perturbations in $A$. $\Phi $ is then
unbounded but $\Phi ^{\prime }$ is bounded (equivalently, $x$ $\simeq \Delta
-\Phi ^{\prime }$ and $y\simeq \Phi ^{\prime \prime }$ are bounded). This
solution is the four wave solution for small $\Lambda $ and $\Delta
=O(\Lambda ).$

In summary, our analysis of the phase equation (\ref{6}) has revealed
several branches of periodic solutions which allows the coexistence of
different periodic states oscillating harmonically in time with similar
periods but exhibiting differing amplitudes. Coexisting periodic states
oscillating harmonically in time were also observed for the laser subject to
optical feedback \cite{al}. Both for the laser subject to injection and the
laser subject to optical feedback, it is the strong phase/amplitude coupling
parameterized by the linewidth enhancement factor $b$ that is the main
mechanism leading to this multiplicity of attractors.

Note that these periodic states appear as isolated branches of solutions.
The traditional way to determine the bifurcation diagram of the long time
regimes is to follow each branch of solutions as a parameter is
progressively changed and carefully determine changes of stabilities between
each states. This method is used both experimentally and numerically but
will fail to find isolated branches of solutions.

Isolated branches of solutions have been found numerically and
experimentally for periodically modulated lasers \cite{glo}. These branches
correspond to different resonances and coexisting periodic states admit
different periods. In the case of a semiconductor laser subject to feedback,
two periodic coexisting attractors were observed experimentally and computed
numerically by Mork {\it et al }\cite{mork}. More recently, De Jagher {\it %
et al }\cite{jhager} reported two coexististing periodic solutions which may
result from the emergence of isolated branches of solutions as described in
this paper. From a practical point of view, it could be interesting to force
the semiconductor laser to operate at these higher intensity states and
investigate possible application in communications or in logic gates.

\section{Discussion}

We have shown analytically that the single mode equations admit multiple
periodic states emerging from limit points in addition to periodic solutions
that appear as the result of a Hopf bifurcation from the steady state. In
general, the spectra of two of the limit cycles might appear very similar if
they are taken at an injection level at which the amplitude saturates and
approaches constant values. However for low injection levels and away from
saturation the spectra can exhibit dramatic differences. This is clearly
demonstrated by the computed spectra shown in Figure 2. In particular small
perturbations in the phase of the field can induce dramatic changes in the
optical spectra. This is precisely the reason that the multiple
instabilities exhibited by the phase of the field can be clearly observed in
the spectra. Such effects include center line suppression, in which the
center line of the laser is almost devoid of energy most of which is carried
by the sidebands at the relaxation frequency. This is a typical feature of a
laser with a large value of the linewidth enhancement factor.

The effect of gain saturation can also be incorporated in the context of the
phase equation formalism. Its main contribution is to shift Hopf bifurcation
and limit points but it would not modify the qualitative properties of the
bifurcation diagram.

We note that the period doubling bifurcation point can also be captured
analytically from Eq. (\ref{6}) by seeking a solution of the form $\Phi
\simeq A\sin (s)+D\cos (s/2).$ If $\Delta =0,$ we obtain

\begin{equation}
\label{17b}\Lambda _{PD}\simeq 0.62 
\end{equation}
which compares well with the prediction of Figure 5a. If $\Delta \neq 0,$ we
find that the period doubling bifurcation point $\Lambda _{PD}(\Delta )$ is
- in parametric form - given by

\begin{equation}
\label{18b}\Lambda _{PD}=\frac{0.32}{J_1(A)},\rm{ }\Delta =\pm 0.32\frac{%
J_0(A)}{J_1(A)}
\end{equation}
where $O(\xi )$ terms have been neglected.

The existence of sustained quasiperiodic oscillations has been suggested
numerically but require a more detailed bifurcation analysis.\ Furthermore,
experimental evidence of quasiperiodic oscillations is still lacking. This
is partly because the laser is not stable at this injection region and
frequently mode hops to a different cavity line. In addition the transition
to the quasiperiodic behavior is a weak phenomenon and it is difficult to
extract the signal from background noise.

In conclusion by taking advantage of the two large parameters $T$ and $b$
that are naturally present in semiconductor lasers, we are able to reduce
the original single mode rate equations into a third order pendulum type
phase equation. This equation then allowed us to predict isolated branching
of solutions which can not be anticipated using traditional continuation
methods.

\section{Appendix: derivation of the phase equation}

We first introduce the decomposition $E=R\exp [i(\Phi -\Omega t)]$ into Eqs.
(\ref{1}) and (\ref{2}) and obtain the following three equations for $A$, $%
\Phi $ and $N$ 
\begin{equation}
\label{a1}R^{\prime }=NR+\eta \cos (\Phi ), 
\end{equation}
\begin{equation}
\label{a2}\Phi ^{\prime }=\Omega -bN-\frac \eta R\sin (\Phi ), 
\end{equation}
\begin{equation}
\label{a3}TN^{\prime }=P-N-P(1+2N)R^2. 
\end{equation}
In Eqs. (\ref{a1})-(\ref{a3}), $T$ is a large parameter that multiplies $%
N^{\prime }.$ We remove this source of singularity by introducing the new
variables $s$, $x$ and $y$ defined by (\ref{4a}). Eqs. (\ref{a1})-(\ref{a3})
then become 
\begin{equation}
\label{a4}y^{\prime }=x(1+b^{-1}y)+\Lambda \cos (\Phi ), 
\end{equation}
\begin{equation}
\label{a5}\Phi ^{\prime }=\Delta -x-\Lambda b^{-1}\frac{\sin (\Phi )}{%
1+b^{-1}y}, 
\end{equation}

\begin{equation}
\label{a6}x^{\prime }=-\xi x-y-2\omega b^{-1}xy-\frac 12b^{-1}y^2-\omega
b^{-2}xy^2. 
\end{equation}
where $\xi ,\Lambda $ and $\Delta $ are dimensionless parameters defined by (%
\ref{5}). Note that $\xi $ is proportional to $\omega $ and that $\omega $
is an $O(T^{-1/2})$ small quantity. Our problem depends on two small
parameters, namely $b^{-1}$ and $\omega .$ The derivation of our simplified
problem will be based on the limit $b^{-1}$ assuming $\omega =O(b^{-1})$.

>From Eqs. (\ref{a5}), we find $x$ as a function $\Phi $

\begin{equation}
\label{a7}x=\Delta -\Phi ^{\prime }-\Lambda b^{-1}\sin (\Phi )+O(b^{-2})
\end{equation}
and from Eq.(\ref{a6}), using (\ref{a7}), we find $y$ as 
\begin{equation}
\label{a7a}y=-\xi x-x^{\prime }+O(b^{-1})=\Phi ^{\prime \prime
}+O(b^{-1},\omega ).
\end{equation}
(Note: the $O(b^{-1})$ correction term in (\ref{a7}) is needed but the $%
O(b^{-1})$ and $O(\omega )$ correction terms in (\ref{a7a}) are not). In
order to derive an equation for $\Phi $ only, we first differentiate Eq. (%
\ref{a6}) once and eliminate $y^{\prime }$ using (\ref{a4}).\ We find 
\begin{equation}
\label{a8}x^{\prime \prime }=-\xi x^{\prime }-\left[ x(1+b^{-1}y)+\Lambda
\cos (\Phi )\right] -b^{-1}y\left( x+\Lambda \cos (\Phi )\right)
+O(b^{-2},b^{-1}\omega ).
\end{equation}
We next use (\ref{a7}) and (\ref{a7a}) and eliminate $x$ and $y$ in (\ref
{a8}).$\;$ This leads to the following third order equation for $\Phi $
given by 
\begin{equation}
\label{a9}
\begin{array}{c}
\Phi ^{\prime \prime \prime }+\Phi ^{\prime }-\Delta -\Lambda \cos (\Phi )
\\ 
+b^{-1}\left[ \Phi ^{\prime \prime }\left( b\xi -2(\Delta -\Phi ^{\prime
})\right) +\Lambda \sin (\Phi )-\Lambda \sin (\Phi )\Phi ^{\prime 2}\right]
+O(b^{-2},b^{-1}\omega )
\end{array}
\end{equation}
where $b\xi =O(b\omega )$ is $O(1).$

\section{Acknowledgments}

This research was supported by the US Air Force Office of Scientific
Research grant AFOSR F49620-94-1-0007,  National Foundation grant
DMS-9625843, the Fonds National de la Recherche Scientifique (Belgium) and
the InterUniversity Attraction Pole of the Belgian government.

\clearpage

\noindent{\bf \large Figure captions}

\bigskip

\parskip=10pt

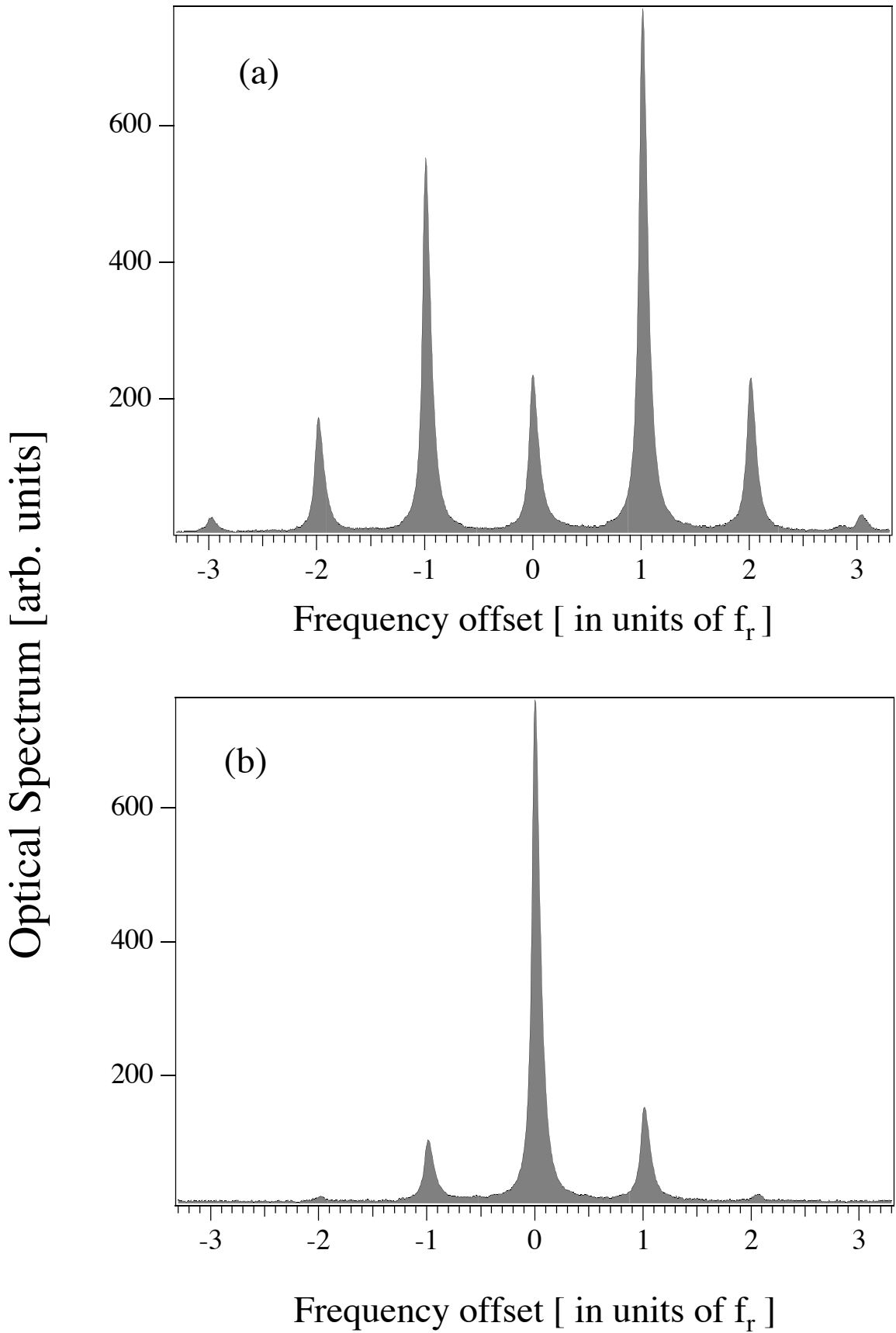
\noindent{\bf Figure 1}
 Experimentally obtained optical spectra of
the laser subject to injection in the neighborhood of the Hopf bifurcation. $%
f_r$ is defined as the free running relaxation oscillation frequency, $%
f_r=\frac 1{2\pi }\sqrt{2P/T}.$

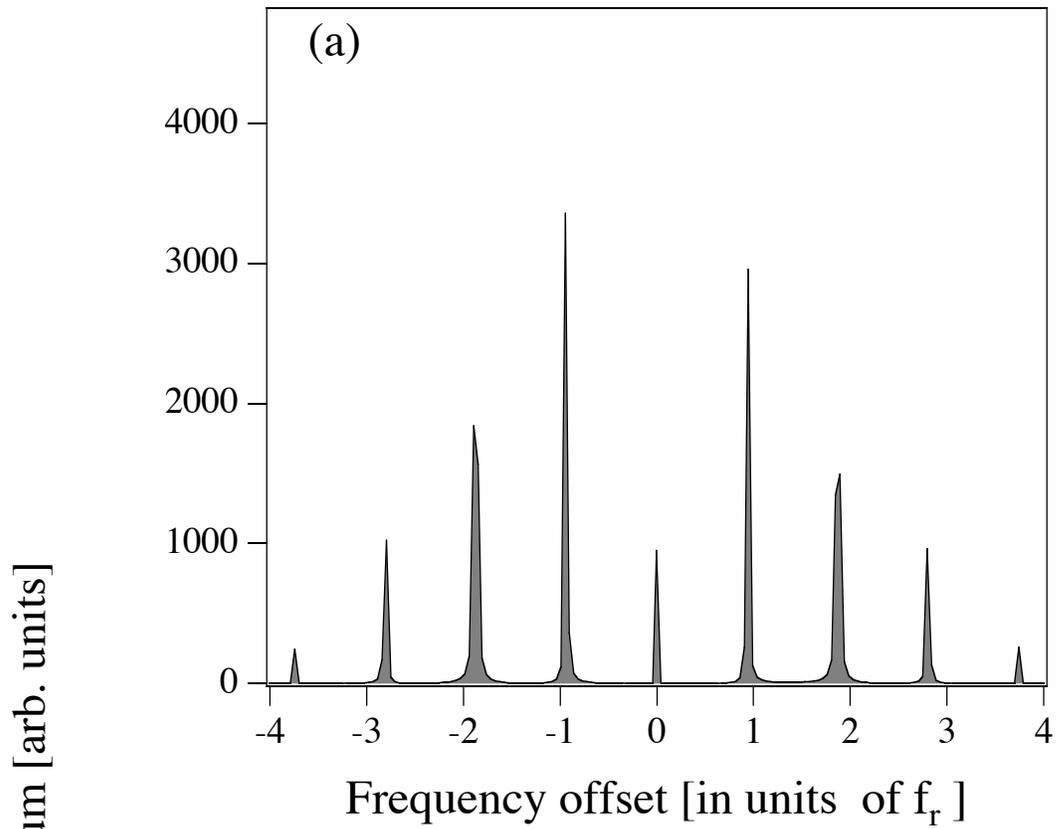
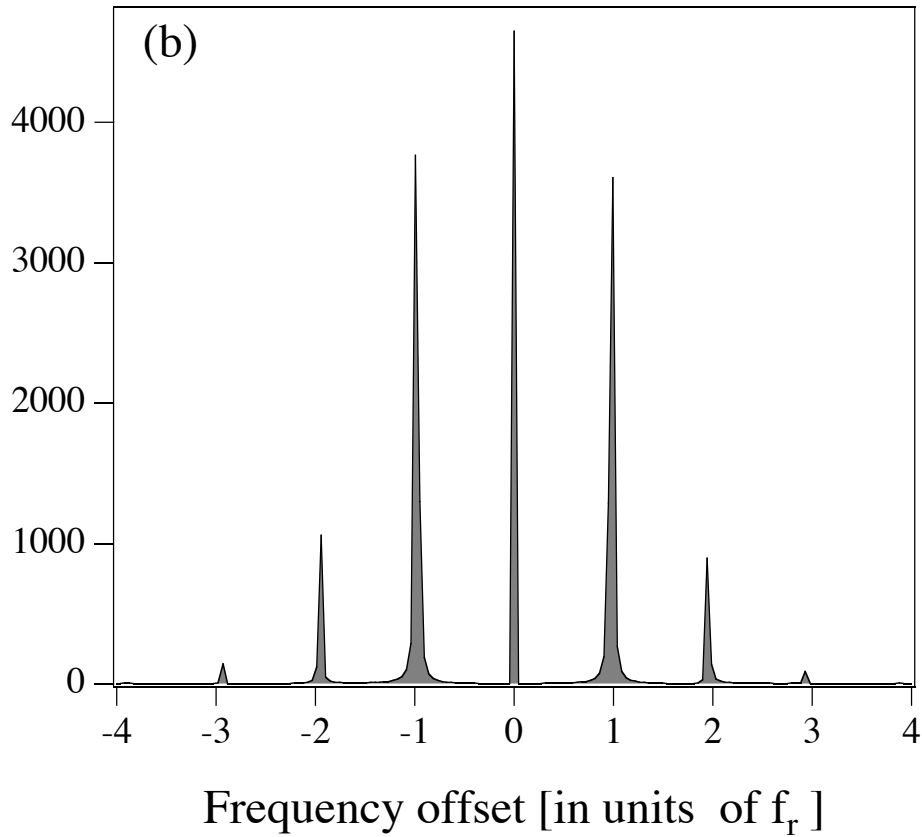
\noindent{\bf Figure 2}
 Numerically computed optical spectra from the
single mode rate equations (\ref{1}) and (\ref{2}) for $T=1000,$ $b=4,$ $%
P=1.0,$ $\eta =0.002$ and $\Omega /\omega =0.1$. The zero of the frequency
scale denotes the frequency of the master laser.

\noindent{\bf Figure 3}
Bifurcation diagram of the extrema of the
deviation of the field from steady state as a function of injection. The
solutions are computed from the rate equations (\ref{1}) and (\ref{2}) using
the same values of the parameters as in Figure 2. The points denoted by A
and B indicate where the spectra in Figures 2a and 2b were numerically computed.

\noindent{\bf Figure 4}
 Bifurcation diagram of the amplitude $A$ of the
periodic solution versus $\Lambda $ constructed form the analytic solutions
of the phase equation for $T=1000,$ $b=10,$ $P=0.5$. Solid lines are for $%
\Delta =0$ and dotted lines for $\Delta =0.01$.

\noindent{\bf Figure 5}
Bifurcation diagram obtained numerically from the
phase equation (\ref{6}) for $T=1000,$ $b=10,$ $P=0.5,$ and zero detuning.%
{\bf \ (a) }shows the extrema of the phase $\Phi $ versus injection
$\Lambda $;
{\bf \ (b)}. shows the extrema of the field versus injection.

\noindent{\bf Figure 6}
Bifurcation diagram obtained numerically from the
phase equation (\ref{6}) for $T=1000,$ $b=10,$ $P=0.5,$ and $\Delta
=0.01. $

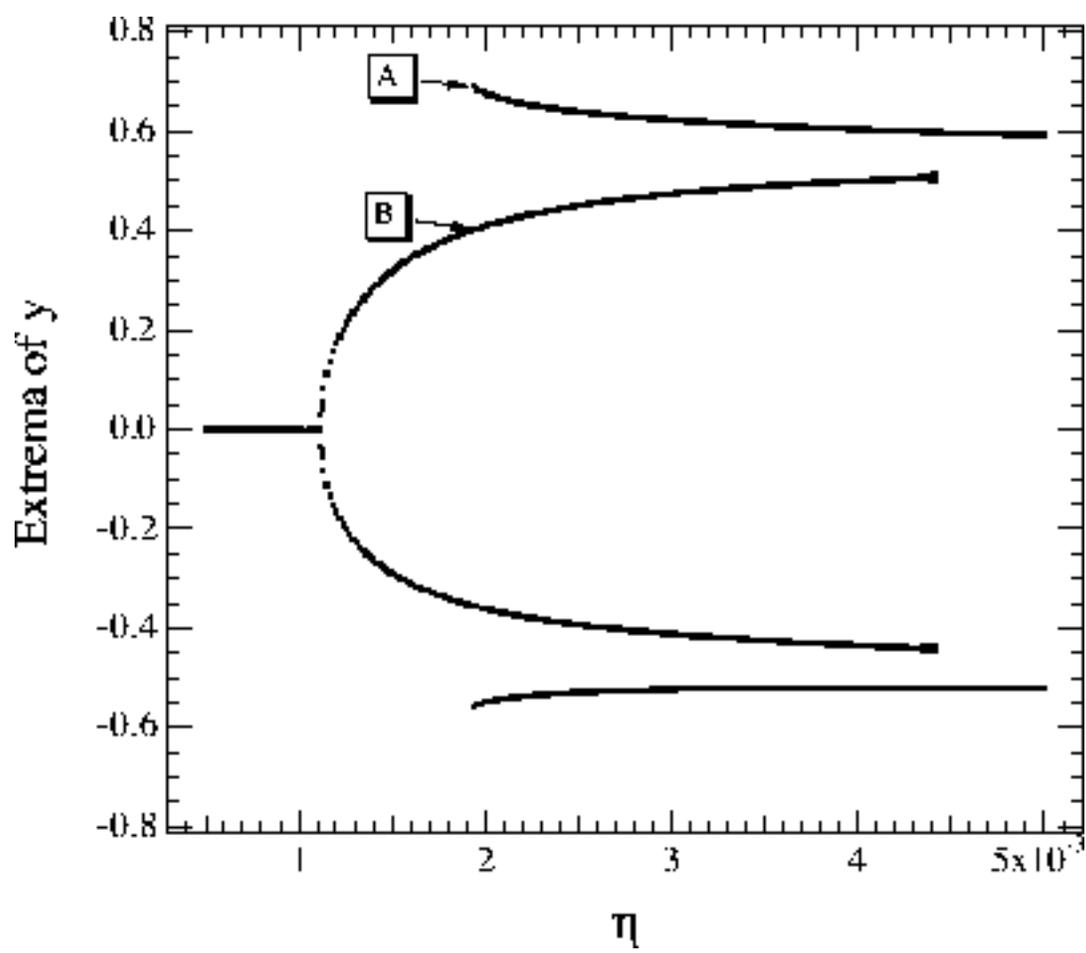
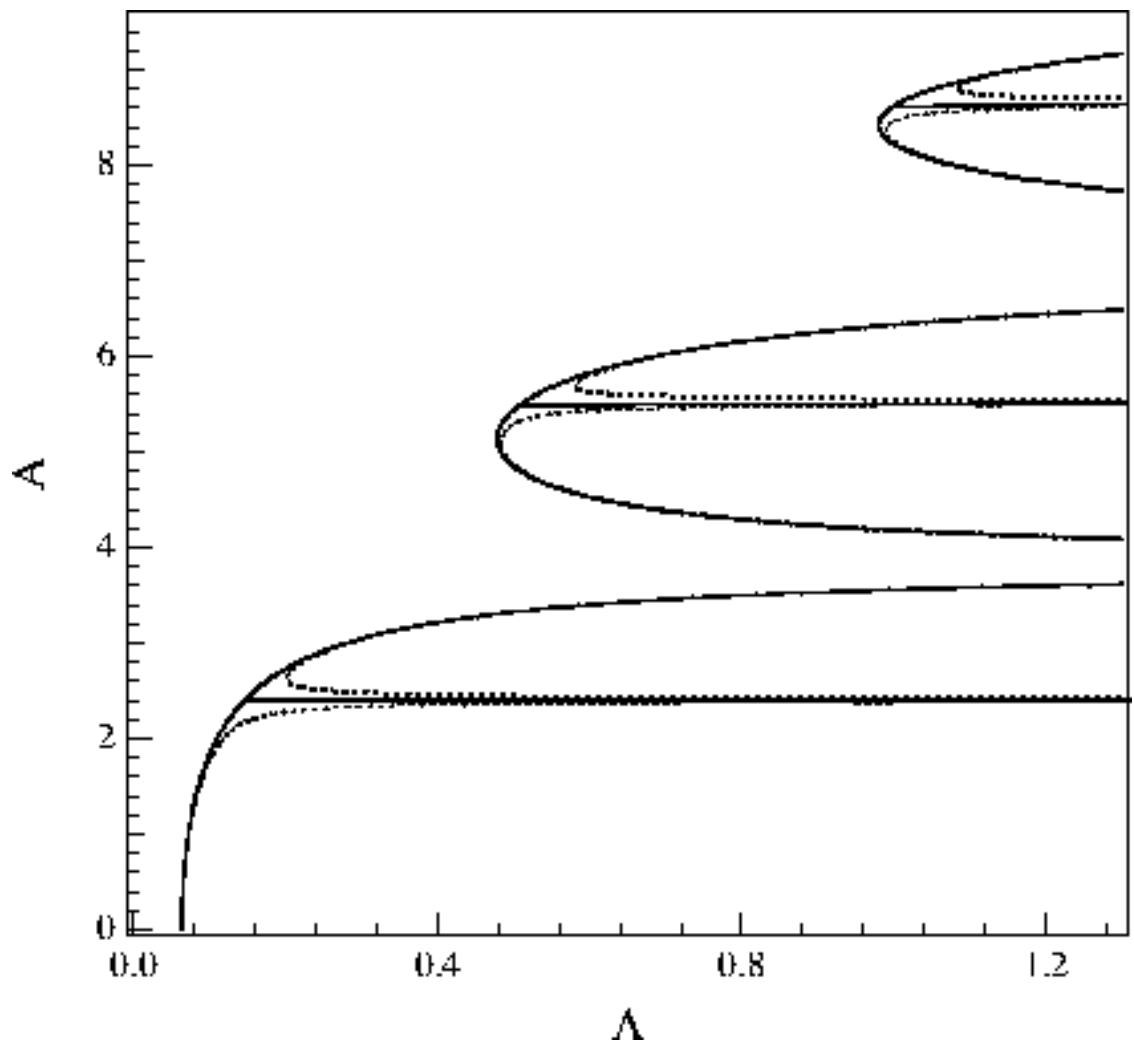
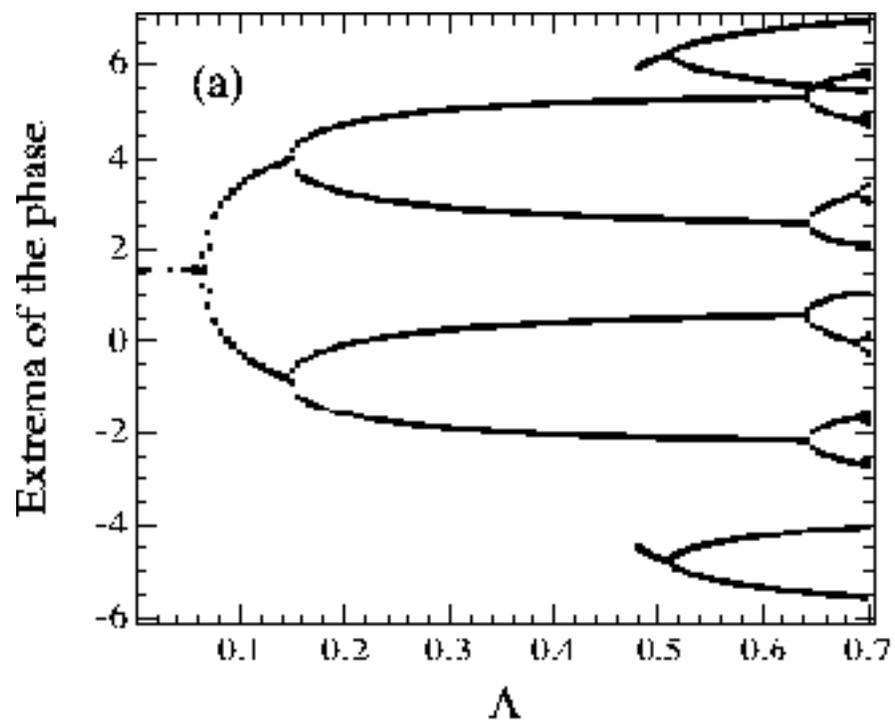
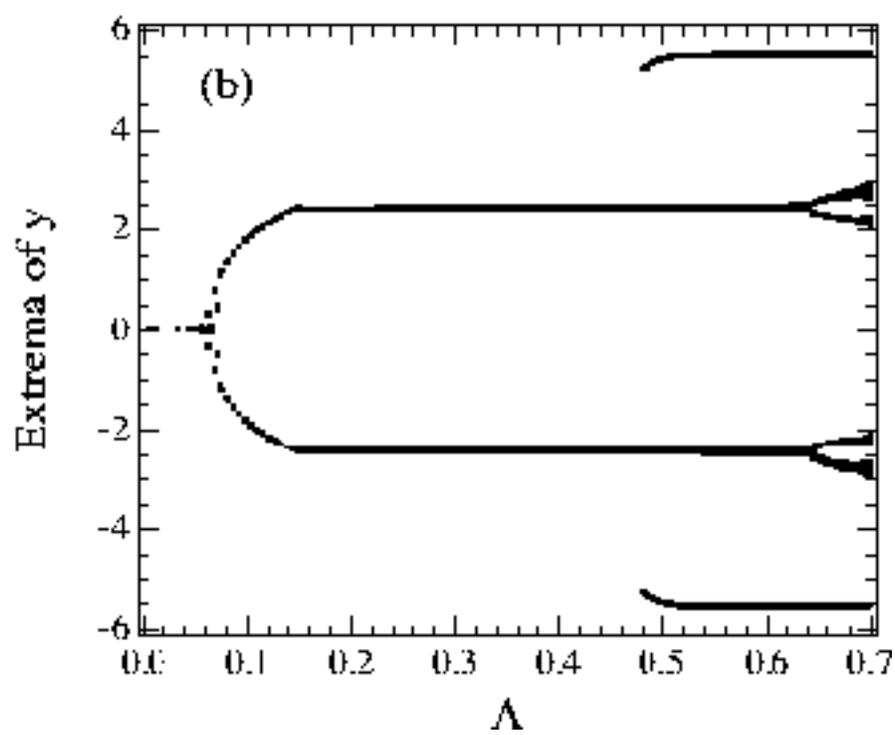
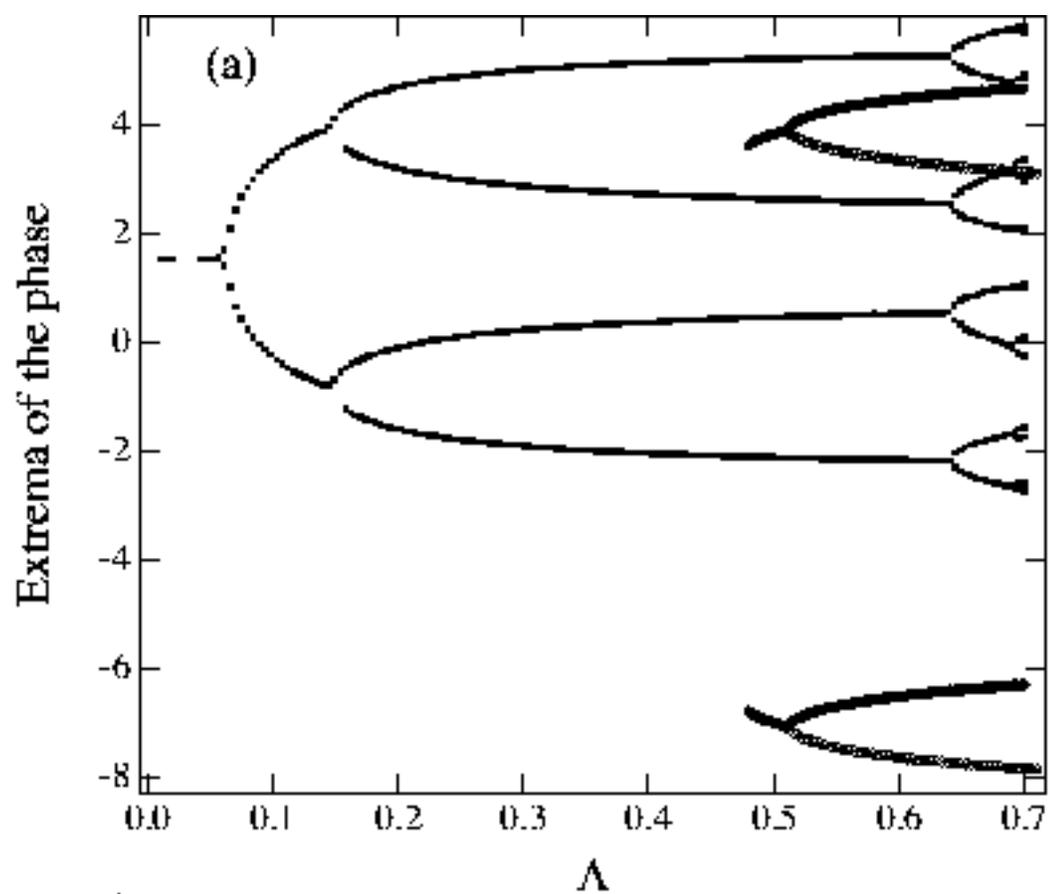
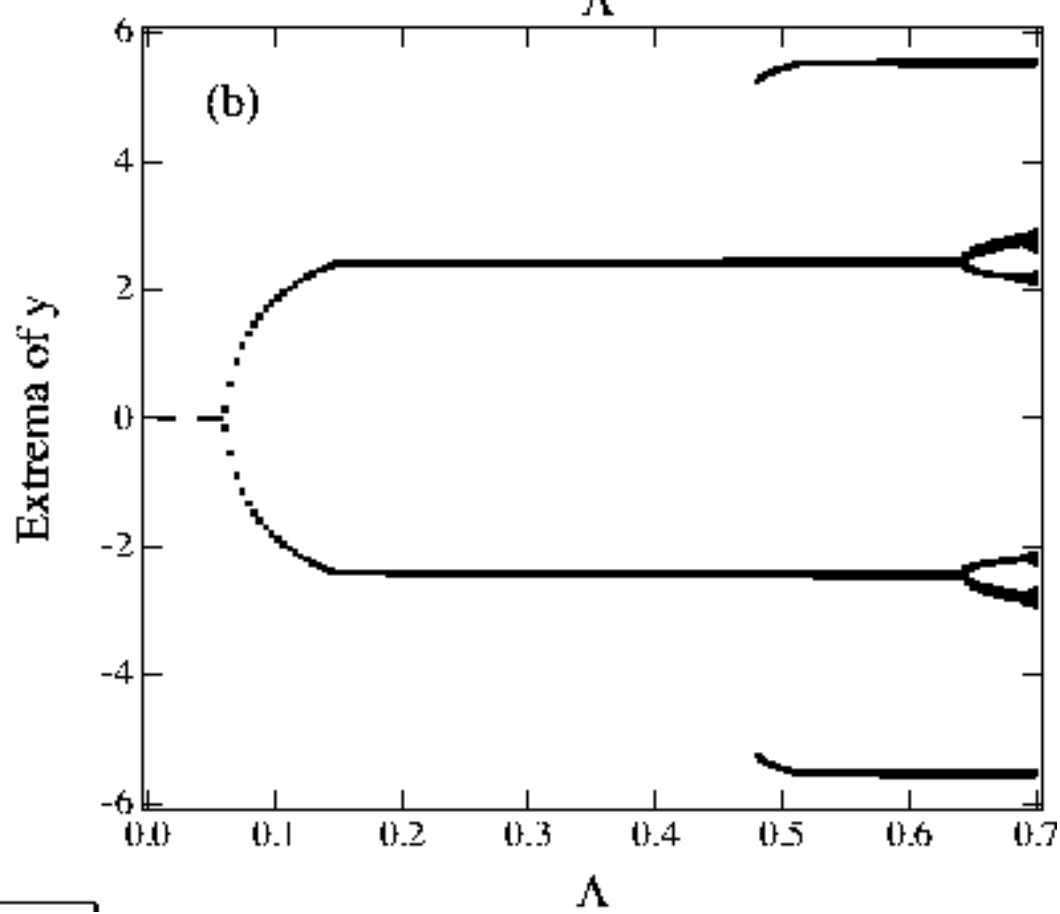
\end{document}